\documentclass[aps,prd,twocolumn]{revtex4}

\usepackage{amsfonts}
\usepackage{amsthm}
\usepackage{amsbsy}
\usepackage{amssymb}
\usepackage{amsmath}
\usepackage{graphicx}
\usepackage{color}

\begin{document}

\date{\today}

% Include your paper's title here

\title{Hawking radiation from ultrashort laser pulse filaments}

\author
{F. Belgiorno,$^{1}$ S.L. Cacciatori,$^{2,3}$ M. Clerici,$^{3}$ V. Gorini,$^{2,3}$\\
G. Ortenzi,$^{4}$ L. Rizzi,$^{3}$ E. Rubino,$^{3}$ V.G. Sala,$^{3}$ D. Faccio$^{3,5\ast}$}

\address{$^{1}$Dipartimento di Fisica, Universit\`a degli Studi di Milano, Via Celoria 16, IT-20133 Milano, Italy\\
$^{2}$INFN sezione di Milano, via Celoria 16, IT-20133 Milano, Italy\\
$^{3}$Dipartimento di Fisica e Matematica, Universit\`a dell'Insubria, Via Valleggio 11, IT-22100 Como, Italy\\
%$^{4}$CNISM e Dipartimento di Fisica e Matematica, Universit\`a dell'Insubria, Via Valleggio 11, IT-22100 Como, Italy\\
$^{4}$Dipartimento di Matematica e Applicazioni, Universit\`a di Milano-Bicocca, Via Cozzi 53, IT-20125 Milano, Italy\\
$^{5}$School of Engineering and Physical Sciences,  Heriot-Watt University, SUPA, Edinburgh EH14 4AS, UK}

\email{ E-mail: d.faccio@hw.ac.uk}

%%%%%%%%%%%%%%%%% END OF PREAMBLE %%%%%%%%%%%%%%%%


%\begin{abstract}
%Event horizons of astrophysical black holes and gravitational analogues have been predicted to excite the quantum vacuum and give rise to the emission of quanta, known as Hawking radiation.   We experimentally create such a gravitational analogue using ultrashort laser pulse filaments and our measurements demonstrate a spontaneous emission of photons that confirms theoretical predictions.\end{abstract}

\maketitle 
\noindent 1) We have shown that our experiments satisfy three of the  conditions listed in Ref.~\cite{unhold_comment}, (time-independence, phase-squeezing, negative energies) and, as argued below,  thermality may not be an essential feature of Hawking emission.\\%The authors of \cite{unhold_comment} start by giving a definition of Hawking radiation which seems to be rather restrictive.\\ % and the actual ensemble of situations to which the definition becomes extremely restricted and cases which are referring to the same physics will be left out.% (even the results in Ref.~\cite{weinfurtner} risk falling short of this definition).  
%This is all the more true when referring to physics that are still not fully understood and are continuously under study, as in the present case.\\
i) %True and exact time-independence cannot be claimed for any known, real-world system. 
Time-independence: the authors of Ref.~\cite{unhold_comment} point out that the surface gravity leads to a characteristic time with respect to which the perturbation must remain stationary.   The surface gravity, as defined in Ref.~\cite{belgiorno}, is of the order of $\kappa=c$/(1 ps). This acceleration gives rise to a variation of the photon velocity of $dv=d(c/n)\sim (c/n^2)dn$. The time scale for this variation is therefore $dv/\kappa\sim 1$ fs, which is more than three orders of magnitude shorter than the time-scale over which nearly stationary, filament-like propagation is observed.  \\
ii) Phase-divergence at the horizon: this has been  shown to occur  \cite{belgiorno}. 
If we trace back in time the outgoing modes in the dispersionless case, we find that they suffer a phase divergence at the horizon with the same logarithmic nature as that pointed out by Hawking in his original work \cite{belgiorno}.
 The phase divergence is maintained also in the dispersive
case, albeit to a limited extent, in agreement with the behaviour in other
dispersive analogues.\\
iii) Thermality: S. Hawking did originally predict a thermal spectrum. Yet it is also known that analogue systems that exhibit dispersion will not necessarily exhibit thermal emission and, even if they do, the temperature may not be related to the surface gravity (e.g. \cite{parentani}).  
It seems reasonable to assume a wider definition that we see based only on the essential ingredients required, namely spontaneous photon or particle excitation due to the presence of an horizon (or even a rapidly forming horizon, see  e.g. \cite{visser1}).\\% where it is implied that the role of the horizon is that of exponentially tearing waves  blocked in it's vicinity.\\
2) The authors of Ref.~\cite{unhold_comment} introduce 
%, apparently ``out of the blue'', 
a condition for which particle creation will occur, namely $\omega_\textrm{frame}^\textrm{pulse}=\omega_\textrm{frame}^\textrm{lab}-\mathbf{v}_\textrm{pulse}\cdot\mathbf{k}$. 
This condition simply implies that the presence of negative frequencies in the comoving reference frame leads to the generation of particles. This condition is so general that it also applies to Hawking radiation.\\
3) The authors comment on the fact that ``there is no exponential tearing (or compaction) by the horizon'' because ``...there is no group velocity horizon'': 
Hawking emission occurs only with negative frequency output modes, i.e. $\omega_\textrm{frame}^\textrm{pulse}<0$ which is clearly related to the existence of an horizon for the {\emph{phase}} velocity, as verified also by our experiments. Regarding group velocity horizons, numerical simulations (to be presented in a future publication) clearly show that mode conversion occurs even in the absence of a group horizon. \\
{ 4) Photon numbers: a calculation of the total photon number emitted by a blackbody, as calculated in the comoving reference frame leads to an estimation that is a few orders of magnitude lower than what is observed in the experiments. The large number of  experimentally measured photons indicates a deviation with respect to the predictions for a static gravitational black hole which, in the different context presented in \cite{PRL} is probably not that surprising. This deviation may either indicate non-stationarity (e.g. related to a non-uniformity of the pulse velocity, as discussed in detail in \cite{njp}) and/or the presence of additional ``boundary'' conditions (e.g. non-trivial input vacuum states \cite{adesso}).
We note that  a full model for emission from superluminal perturbations has been developed and gives a scaling factor $\delta n^2$ but, contrary to what is claimed in \cite{unhold_comment}, does give the same order of magnitude for the emitted photon numbers as in the measurements. However this model was excluded as it is a distinct effect from Hawking emission and does not capture the main spectral features of \cite{PRL}. Finally, two very recent non-perturbative models for Hawking emission that include dispersion both  predict photon numbers that are of the same order of magnitude \cite{scott} or higher than in the measurements \cite{guerreiro}. }
%Our experiments therefore satsify three out of the four conditions listed (thermality, time-independence, phase-squeezing, negative energies) and we do not consider thermality as being an essential feature of Hawking radiation.
%We note that, regarding the available models, dispersion does need to be accounted for, ab-initio, in a full 4D model and this will hopefully lead to a quantitative prediction of the number of emitted photons. We limit ourselves here to noting that the small photon counts typically expected in the visible-near-infrared region are due solely to the exponential cut-off at high frequencies associated to the blackbody spectrum (if one makes this assumption). However, if one removes this exponential cut-off and replaces it e.g. with a polynomial or even flat-spectral dependence, it is easy to see that the situation will be very different and very high photon counts become possible. Further work is certainly needed in this direction and is currently underway.\\
%Nevertheless, we believe that all the necessary ingredients for Hawking-like emission are in place and do explain the results in \cite{PRL}.\\

\end{document}